\documentclass[twocolumn, superscriptaddress,pra,aps,showpacs]{revtex4-1}
\usepackage{amsmath,mathrsfs,amsbsy,color,graphicx,bm,amsthm,amsfonts,array,multirow}
\usepackage[english]{babel}
\usepackage[latin1]{inputenc}
\usepackage{graphicx}
\usepackage{braket}
\usepackage{tabularx}
\usepackage{afterpage}
\usepackage{mathdots}

\newtheorem{observation}{Observation}

\newtheorem{theorem}{Theorem}

\begin{document}

\title{The multipartite entanglement classes of a multiport beam-splitter}

\author{F. E. S. Steinhoff} 
\email{steinhofffrank@gmail.com}
\affiliation{Faculdade de Engenharia, Universidade Federal de Mato Grosso, 78060-900 V\'arzea Grande, Mato Grosso, Brazil}


\date{\today}

\begin{abstract} 
The states generated by a multiport beam-splitter usually display genuine multipartite entanglement between the many spatial modes. Here we investigate the different classes of multipartite entangled states that arise in this practical situation, working within the paradigm of Stochastic Local Operations with Classical Communication. We highlight three scenarios, one where the multipartite entanglement classes follow a total number hierarchy, another where the various classes follow a nonclassicality degree hierarchy and a third one that is a combination of the previous two. Moreover, the multipartite entanglement of higher-dimensional versions of Dicke states relate naturally to our results.      
\end{abstract}

\maketitle

\section{Introduction}

Multipartite entanglement is a fundamental property of quantum systems, enabling the success of tasks that would be unthinkable classically \cite{nielsen,multientang, gtreview,cris}. It has deep connections to many-body physics \cite{manybody,mps} as well as foundational aspects of quantum theory \cite{nonlocality}. In the paradigm of Stochastic Local Operations with Classical Communication (SLOCC), there are inequivalent classes of multipartite entangled states, each class having radically distinct physical and informational properties \cite{wghz}. A considerable number of criteria has been developed to address the problem of SLOCC classification \cite{miyake,verstraete,lamata,dafa,gour,elos}, which is known to be highly non-trivial from a computational point of view.  

In the realm of quantum optics, a multiport beam-splitter (MBS) can be employed for the generation of certain specific multipartite entangled states \cite{mbs,mbs2}. Our approach in the present work, however, is to consider typical output states of a MBS and to then determine the SLOCC classes that arise in this practical scenario. These multimode entangled states are the result of the action of the MBS on a single-mode input state, revealing nonclassical features of this state \cite{nonc,nonc2,nonc3,nonc4}. 

The manuscript is divided as follows: in Section II, we give a brief review of the problem of SLOCC classification in general, as well as the main features of a multiport beam-splitter. In Section III, we show various results concerning the multipartite entanglement classes of the states generated by a multiport beam-splitter. We consider three paradigmatic scenarios according to the input state: (a) input states that are finite superpositions of number states and (b) input states that are finite superpositions of coherent states and (c) input states that are finite superpositions of both number and coherent states. This restriction to finite superpositions comes from the observations in \cite{infinite}, where it is shown that even in the bipartite case there exist states with an infinite degree of entanglement that cannot be connected through SLOCC. We show that the multipartite entanglement classes of scenario (a) are governed by a hierarchy based on the highest total number of photons in the state, while the classes in scenario (b) follow a hierarchy based on the so-called nonclassicality rank \cite{nonc,nonc2}. Interestingly, some superpositions of qudit Dicke states \cite{ds1,ds2,ds3} and a recent generalization of these \cite{ds4} fall in the classification scheme of scenario (a). In Section IV we compare the three types of states arising from scenarios (a), (b) and (c), showing that each scenario constitute a multipartite entanglement class of its own. Finally in Section V, we discuss the conclusions of the results as well as perspectives on future problems. 

\section{Preliminaries}

\subsection{Multipartite entanglement classes}

We consider a state space $\mathcal{H}$ composed of $m$ subsystems $\mathcal{H}_1,\mathcal{H}_2,\ldots,\mathcal{H}_m$, i.e., $\mathcal{H}=\bigotimes_{k=1}^m\mathcal{H}_k$. Let $dim(\mathcal{H}_k)=d_k$ and let $\{|e_1,e_2,\ldots,e_m\rangle\}$ denote the computational basis of $\mathcal{H}$, with $e_k=0,1,\ldots,d_k-1$. An arbitrary state $|\Psi\rangle\in\mathcal{H}$ is written in the computational basis as
\begin{eqnarray*}
    |\Psi\rangle=\sum_{e_k=0}^{d_k-1}\Psi_{e_1e_2\ldots e_m}|e_1,e_2,\ldots,e_m\rangle
\end{eqnarray*}
where the scalars $\Psi_{e_1e_2\ldots e_m}$ constitute the so-called coefficient tensor of the state $|\Psi\rangle$. For the purposes of this work, we can  assume $d_1=d_2=\ldots=d_m=d$. We can arrange the coefficients $\Psi_{e_1e_2\ldots e_m}$ in a matrix $\mathcal{M}_{|\Psi\rangle}=(\Psi_{e_1,e_2\ldots e_m})$, called the coefficient matrix of $|\Psi\rangle$, where the rows are indexed by the values $e_1=0,1,\ldots,d-1$, while the columns are indexed by the remaining values $e_2\ldots e_m$ in lexicographic order. We can then arrange $\mathcal{M}_{|\Psi\rangle}$ as 
\begin{eqnarray*}
    \mathcal{M}_{|\Psi\rangle}=\left(\begin{array}{c|c|c|c}
         {M_0}&{M_1}&{\ldots}&{M_{d^{m-2}-1}} 
    \end{array}\right)
\end{eqnarray*}
where each submatrix $M_k$ is a $d\times d$ block. For the bipartite case $m=2$ the coefficient matrix is a proper matrix
\begin{eqnarray*}
     \mathcal{M}_{|\Psi\rangle}=\left(\begin{array}{cccc}
         {\Psi_{0,0}}&{\Psi_{0,1}}&{\ldots}&{\Psi_{0,d-1}}\\
         {\Psi_{1,0}}&{\Psi_{1,1}}&{\ldots}&{\Psi_{1,d-1}}\\
         {\vdots}&{\vdots}&{}&{\vdots}\\
        {\Psi_{d-1,0}}&{\Psi_{d-1,1}}&{\ldots}&{\Psi_{d-1,d-1}}
    \end{array}\right)
\end{eqnarray*}
The coefficient matrix for the case $d=2,m=3$ (three qubits) is given by
\begin{eqnarray*}
     \mathcal{M}_{|\Psi\rangle}=\left(\begin{array}{cc|cc}
         {\Psi_{0,00}}&{\Psi_{0,01}}&{\Psi_{0,10}}&{\Psi_{0,11}}\\
         {\Psi_{1,00}}&{\Psi_{1,01}}&{\Psi_{1,10}}&{\Psi_{1,11}}
    \end{array}\right)
\end{eqnarray*}
In general, we say that a state $\rho$ is SLOCC-equivalent to $\rho'$ (notation: $\rho\sim\rho'$) if one can be obtained from the other by the sole use of SLOCC procedures; otherwise, we call the states SLOCC-inequivalent. For pure states, the problem simplifies significantly \cite{multientang}:
\begin{observation}
    The $m$-partite pure states $|\Psi\rangle$ and $|\Phi\rangle$ are SLOCC-equivalent if and only if there exists Invertible Local Operators (ILOs) $A_k$ such that
    \begin{eqnarray*}
        |\Phi\rangle = A_1\otimes A_2\otimes\ldots\otimes A_m|\Psi\rangle
    \end{eqnarray*}
\end{observation}
Besides entanglement, ILOs play an important role in hidden nonlocality \cite{lf,lf2}, where they are known as Local Filters. The action of ILOs on a state $|\Psi\rangle$ induces linear operations on the rows and columns of the coefficient matrix $\mathcal{M}_{|\Psi\rangle}$. For the bipartite case $m=2$, the rank of $\mathcal{M}_{|\Psi\rangle}$, called the Schmidt rank of $|\Psi\rangle$, is invariant under ILOs and thus two bipartite pure states are SLOCC-equivalent if and only if they have the same (finite) Schmidt rank \cite{elos,vidal}. For $m>3$, we say that a state is genuinely mutipartite entangled if it is entangled with respect to any bipartition of the state space, i.e., if the Schmidt rank of any bipartition is greater than $1$. 

The three-qubit case $d=2,m=3$ reveals that there are two SLOCC-inequivalent classes of genuinely entangled states \cite{wghz}: the GHZ class, with representative $|GHZ\rangle=2^{-1/2}(|000\rangle+|111\rangle)$ and the W class, with representative $|W\rangle=3^{-1/2}(|001\rangle+|010\rangle+|100\rangle)$. For the four-qubit case $d=2,m=4$ it is known that there exists an infinite number of SLOCC-inequivalent genuinely entangled states and it is then more practical to classify states in terms of a finite number of families of SLOCC-equivalent classes \cite{verstraete}.  

For the construction of some of the ILOs connecting SLOCC-equivalent states, we use the ideas of \cite{elos}, where ILOs are decomposed as finite sequences of Elementary Local Operations (ELOs). It is possible to then employ a multipartite version of the Gauss-Jordan Elimination procedure on the coefficient matrix of a state in order to map it into the coefficient matrix of a suitable state that is representative of this class. Other multipartite entnglement classification schemes based on the coefficient matrix are found in \cite{lamata,dafa,cm,cm2,cm3}.

\subsection{States generated by a Multiport Beam-Splitter}

A Multiport Beam-Splitter (MBS) is an optical device that implements linear operations on the creation and annihilation operators on $m$ spatial modes \cite{puri}. Specifically, if we write a vector $\mathbf{a}^{\dagger}=(a_1^{\dagger},a_2^{\dagger},\ldots,a_m^{\dagger})$, then the effect of the MBS is to perform the map $\mathbf{a}^{\dagger}\rightarrow S\mathbf{a}^{\dagger}$, where $S$, the so-called scattering matrix, is an element of $\mathcal{SU}(m)$; a similar transformation takes place for the annihilation operators. Alternatively, one can see the effect of the MBS as the unitary action $a^{\dagger}_k\rightarrow U_Sa_k^{\dagger}U_S^{\dagger}$, for $k=1,\ldots,m$.

We will be mainly interested in input states of the form $|\psi\rangle\otimes|0,0,\ldots,0\rangle$, where all modes but the first one are equal to the vacuum state. The reason for this restriction is to avoid interference effects between different modes, which can complicate considerably the problem of entanglement classification \cite{futurework,goldbergjames,multentanggen}. Moreover, the MBS is a passive physical device and any quantum correlation present at its exit can be traced back to the quantum correlations of the single-mode state $|\psi\rangle$ entering the MBS \cite{nonc,nonc2,nonc3,nonc4}. With this restriction in mind, we show in Appendix A that a balanced MBS on the first mode, in the sense that 
\begin{eqnarray*}
    a_1^{\dagger}\rightarrow U_Sa_1^{\dagger}U^{\dagger}_S=\frac{a_1^{\dagger}+a_2^{\dagger}+\ldots+a_m^{\dagger}}{\sqrt{m}}
\end{eqnarray*} 
is sufficient for the SLOCC classification of output states of any MBS; thus, without loss of generality, in what follows we only consider the action of a balanced MBS. 

A number state on the first mode is given by
\begin{eqnarray*}
    |n,0,\ldots,0\rangle=\frac{(a_1^{\dagger})^{n}}{\sqrt{n!}}|0,0,\ldots,0\rangle 
\end{eqnarray*}
The output state of the MBS $|\Psi_n\rangle=U_S|n,0,\ldots,0\rangle$ is thus 
\begin{eqnarray*}
    |\Psi_n\rangle=\frac{1}{\sqrt{n!}}\left(\frac{a_1^{\dagger}+a_2^{\dagger}+\ldots+a_m^{\dagger}}{\sqrt{m}}\right)^n|0,0,\ldots,0\rangle\\
    =\sum_{n_1+\ldots+n_m=n}{n\choose{n_1,\ldots,n_m}}\frac{(a_1^{\dagger})^{n_1}\ldots(a_m^{\dagger})^{n_m}}{\sqrt{n!m^n}}|0,\ldots,0\rangle
\end{eqnarray*}
where the summation runs through all the possible non-negative integer values $n_1,\ldots,n_m$ satisfying the constraint $n_1+\ldots+n_m=n$. Some simple algebraic manipulation results in 
\begin{eqnarray*}
    |\Psi_n\rangle=\frac{1}{m^{n/2}}\sum_{n_1+\ldots +n_m=n}\sqrt{{n}\choose{n_1,\ldots,n_m}}|n_1\ldots n_m\rangle.
\end{eqnarray*}
This state is symmetrical with respect to any permutation of the modes.

Another interesting type of input state is the coherent state 
\begin{eqnarray*}
    |\alpha,0,\ldots,0\rangle=D_1(\alpha)|0,0,\ldots,0\rangle
\end{eqnarray*}
where $D_1(\alpha)=e^{\alpha a_1^{\dagger}-\alpha^* a_1}$ is the displacement operator. The output state $U_S|\alpha,0,\ldots,0\rangle$ then reads
\begin{eqnarray*}
    |\eta_{\alpha}\rangle=|\alpha/\sqrt{m},\alpha/\sqrt{m},\ldots,\alpha/\sqrt{m}\rangle,
\end{eqnarray*}
since $U_SD_1(\alpha)U_S^{\dagger}=D_1(\alpha/\sqrt{m})D_2(\alpha/\sqrt{m})\ldots D_m(\alpha/\sqrt{m})$.

\section{Results}

\subsection{Superpositions of number states as input}

We consider a $m$-port balanced beam-splitter and an input state on the first mode given by
\begin{eqnarray*}
    |\psi\rangle\otimes|0,\ldots,0\rangle=\sum_{n=0}^Nc_n|n,0,\ldots,0\rangle,
\end{eqnarray*}
representing an arbitrary finite superposition of number states. The value $N$ represents the highest total number of photons/energy. A finite superposition of number states can be obtained from an infinite one via truncation methods \cite{truncation,truncation2,cutoff}. The resulting output state of the MBS is then 
\begin{eqnarray*}
    |\Psi\rangle=U_S|\psi\rangle\otimes|0,\ldots,0\rangle=\sum_{n=0}^Nc_n|\Psi_n\rangle,
\end{eqnarray*}
where, according to the previous section, 
\begin{eqnarray*}
    |\Psi_n\rangle=\sum_{n_1+\ldots +n_m=n}C^n_{n_1\ldots n_m}|n_1\ldots n_m\rangle;\\
    C^n_{n_1\ldots n_m}=\frac{1}{m^{n/2}}\sqrt{{n}\choose{n_1,\ldots,n_m}}
\end{eqnarray*}
We now show that $|\Psi\rangle$ can be brought to a more convenient form $|\Phi\rangle =\sum_{n=0}^Nh_n|\Phi_n\rangle$, where $h_n=c_n\sqrt{\frac{n!}{m^n}}$ and we define the unnormalized states \footnote{An unnormalized state $|\nu\rangle$ is trivially SLOCC-equivalent to its normalized version via the ILO $\frac{1}{\sqrt{\langle\nu|\nu\rangle}}\mathcal{I}$.}
\begin{eqnarray*}
    |\Phi_n\rangle=\sum_{n_1+\ldots +n_m=n}|n_1\ldots n_m\rangle
\end{eqnarray*}
which we call uniform states. Some superpositions of these uniform states were considered in \cite{elos}, based on the related work \cite{becv} and having interesting properties such as coefficient matrices with special structures. Moreover, some of the so-called qudit Dicke states \cite{ds1,ds2,ds3}, as well as the recent spin-$s$ Dicke states \cite{ds4} can be seen as specific superpositions of the uniform states $|\Phi_n\rangle$, i.e., are special cases of states in the form $|\Phi\rangle =\sum_{n=0}^Nh_n|\Phi_n\rangle$ and hence their SLOCC classification can be obtained within our framework.    
\begin{observation}
The output state $|\Psi\rangle$ is SLOCC-equivalent to 
$|\Phi\rangle$. In particular, $|\Psi_n\rangle\sim|\Phi_n\rangle$ for any value of $n$.
\end{observation}
{\it Proof:} Given the following invertible operation on mode $k$ 
\begin{eqnarray*}
    R_k=\sum_{n_k=0}^N\sqrt{n_k!}|n_k\rangle\langle n_k|,
\end{eqnarray*}
we notice first that
\begin{eqnarray*}
   \bigotimes_{k=1}^mR_k|\Psi_n\rangle&=&\sum_{n_1+\ldots +n_m=n}C^n_{n_1\ldots n_m}\bigotimes_{k=1}^mR_k|n_1\ldots n_m\rangle\\
    &=&\sum_{n_1+\ldots +n_m=n}C^n_{n_1\ldots n_m}\sqrt{n_1!\ldots n_m!}|n_1\ldots n_m\rangle\\
    &=&\sqrt{\frac{n!}{m^n}}\sum_{n_1+\ldots +n_m=n}|n_1\ldots n_m\rangle=\sqrt{\frac{n!}{m^n}}|\Phi_n\rangle
\end{eqnarray*}
We thus conclude that
\begin{eqnarray*}
    \bigotimes_{k=1}^mR_k|\Psi\rangle=\sum_{n=0}^Nc_n\bigotimes_{k=1}^mR_k|\Psi_n\rangle=\sum_{n=0}^Nh_n|\Phi_n\rangle=|\Phi\rangle
\end{eqnarray*}
where $h_n=c_n\sqrt{\frac{n!}{m^n}}$.  $\Box$

We are now ready to show that the output state $|\Psi\rangle=\sum_{n=0}^Nc_n|\Psi_n\rangle$ is SLOCC-equivalent to $|\Psi_N\rangle$, i.e., the term with highest number of photons:  
\begin{observation}
The uniform state $|\Phi\rangle$ is SLOCC-equivalent to $|\Phi_N\rangle$. Moreover, we have the following chain of SLOCC-equivalences:
\begin{eqnarray*}
    |\Psi\rangle\sim|\Phi\rangle\sim|\Phi_N\rangle\sim|\Psi_N\rangle.
\end{eqnarray*}
\end{observation}
{\it Proof:} In what follows, $\overline{x}$ denotes the number $x/(N+1)$ in base $N+1$. For example, if $N=3$, then $\overline{56}=32$, since $56/4=14$ and $14$ equals $32=3.4^1+2.4^0$ in base $4$. The state $|\Phi\rangle$ has coefficient matrix given by $\mathcal{M}_{|\Phi\rangle}=(\Phi_{n_1,n_2\ldots n_m})$. We then write the array $n_2\ldots n_m$, which gives the columns of $\mathcal{M}_{|\Phi\rangle}$, as $\overline{n_2\ldots n_m}$. The coefficient matrix of $|\Phi\rangle$ can then be written as  
\begin{eqnarray*}
    \mathcal{M}_{|\Psi\rangle}=\left(\begin{array}{c|c|c|c|c}
         {M_{\overline{0}}}&{M_{\overline{1}}}&{M_{\overline{2}}}&{\ldots}&{M_{\overline{d^{m-2}-1}}} 
    \end{array}\right)
\end{eqnarray*}
where the submatrices $M_{\overline{k}}$ are indexed according to their first column $k$ with decimal expression $n_2n_3\ldots n_m$; notice that $k$ is then a multiple of $N+1$. This notation is to take into account the repetitions of submatrices that occur for $m>3$, as well as identifying null submatrices and allowing possible future generalizations for the infinite superposition case. Moreover, let $||\vec{k}||=n_2+n_3+\ldots+n_m$. The submatrices $M_{\overline{k}}$ are Hankel matrices \cite{horn} of a special form,  
\begin{eqnarray*}
    M_{\overline{0}}=\left(\begin{array}{cccccc}
         {h_0}&{h_1}&{h_2}&{h_3}&{\ldots}&{h_N}  \\
         {h_1}&{h_2}&{h_3}&{\ldots}&{h_N}&{0}  \\
         {h_2}&{h_3}&{}&{\ \iddots}&{0}&{\vdots}  \\
         {h_3}&{\vdots}&{\iddots}&{\iddots}&{}&{}  \\
         {\vdots}&{h_N}&{0}&{}&{}&{\vdots}  \\
         {h_N}&{0}&{\ldots}&{}&{\ldots}&{0}
    \end{array}\right), 
\end{eqnarray*}
\begin{eqnarray*}
M_{\overline{k}}=\left(\begin{array}{cccccc|c}
         {h_1}&{h_2}&{h_3}&{h_4}&{\ldots}&{h_N}&{0}  \\
         {h_2}&{h_3}&{h_4}&{\ldots}&{h_N}&{0}&{0}  \\
         {h_3}&{h_4}&{}&{\ \iddots}&{0}&{\vdots}&{\vdots}  \\
         {h_4}&{\vdots}&{\iddots}&{\iddots}&{}&{}&{\vdots}  \\
         {\vdots}&{h_N}&{0}&{}&{}&{\vdots}&{0}  \\
         {h_N}&{0}&{\ldots}&{}&{\ldots}&{0}&{0}  \\
 \hline  {0}&{0}&{\ldots}&{}&{\ldots}&{0}&{0}
    \end{array}\right) \\ ||\vec{k}||=1,
\end{eqnarray*}
\begin{eqnarray*}
M_{\overline{k}}=\left(\begin{array}{cccccc|cc}
         {h_2}&{h_3}&{h_4}&{h_5}&{\ldots}&{h_N}&{0}&{0}  \\
         {h_3}&{h_4}&{h_5}&{\ldots}&{h_N}&{0}&{0}&{0}  \\
         {h_4}&{h_5}&{}&{\ \iddots}&{0}&{\vdots}&{\vdots}&{\vdots}  \\
         {h_5}&{\vdots}&{\iddots}&{\iddots}&{}&{}&{\vdots}&{\vdots}  \\
         {\vdots}&{h_N}&{0}&{}&{}&{\vdots}&{0}&{0}  \\
         {h_N}&{0}&{\ldots}&{}&{\ldots}&{0}&{0}&{0}  \\
  \hline {0}&{0}&{\ldots}&{}&{\ldots}&{0}&{0}&{0}\\
         {0}&{0}&{\ldots}&{}&{\ldots}&{0}&{0}&{0}
    \end{array}\right), \\ ||\vec{k}||=2,
\end{eqnarray*}
and an arbitrary submatrix with $||\vec{k}||=n$ is given by
\begin{eqnarray*}
    M_{\overline{k}}=\left(\begin{array}{cccccc|ccc}
         {h_n}&{h_{n+1}}&{h_{n+2}}&{h_{n+3}}&{\ldots}&{h_N}&{0}&{\ldots}&{0}  \\
         {h_{n+1}}&{h_{n+2}}&{h_{n+3}}&{\ldots}&{h_N}&{0}&{0}&{\ldots}&{0}  \\
         {h_{n+2}}&{h_{n+3}}&{}&{\ \iddots}&{0}&{\vdots}&{\vdots}&{}&{\vdots}  \\
         {h_{n+3}}&{\vdots}&{\iddots}&{\iddots}&{}&{}&{}&{}&{}  \\
         {\vdots}&{h_N}&{0}&{}&{}&{\vdots}&{\vdots}&{}&{\vdots}  \\
         {h_N}&{0}&{\ldots}&{}&{\ldots}&{0}&{0}&{\ldots}&{0}  \\
 \hline  {0}&{0}&{\ldots}&{}&{\ldots}&{0}&{0}&{\ldots}&{0}\\
         {\vdots}&{\vdots}&{}&{}&{}&{\vdots}&{\vdots}&{}&{\vdots}\\
         {0}&{0}&{\ldots}&{}&{\ldots}&{0}&{0}&{\ldots}&{0}
    \end{array}\right).
\end{eqnarray*}
Notice that if $||\vec{k}||>N$, the submatrix $M_{\overline{k}}$ is a null matrix. Let $A=A^{(N-1)}A^{(N-2)}\ldots A^{(1)}A^{(0)}$, where $A^{(k)}=\mathcal{I}+\sum_{n=k}^{N-1}\lambda_n|n\rangle\langle N-k|$, with $\mathcal{I}$ denoting the identity operator and $\lambda_n=-h_n/h_N$. The matrices representing these operators in number basis are given by
\begin{eqnarray*}
    A^{(0)}=\left(\begin{array}{ccccc}
         {1}&{}&{}&{}&{\lambda_0}  \\
         {}&{1}&{}&{}&{\lambda_1}  \\
         {}&{}&{\ddots}&{}&{\vdots}  \\
         {}&{}&{}&{1}&{\lambda_{N-1}}  \\
         {}&{}&{}&{}&{1}  
    \end{array}\right), 
\end{eqnarray*}
\begin{eqnarray*}
    A^{(1)}=\left(\begin{array}{cccccc}
         {1}&{}&{}&{}&{\lambda_1}&{}  \\
         {}&{1}&{}&{}&{\lambda_2}&{}  \\
         {}&{}&{\ddots}&{}&{\vdots}&{}  \\
         {}&{}&{}&{1}&{\lambda_{N-1}}&{}  \\
         {}&{}&{}&{}&{1}&{}  \\
         {}&{}&{}&{}&{}&{1}  
    \end{array}\right)
\end{eqnarray*}
and, in general,
\begin{eqnarray*}
    A^{(k)}=\left(\begin{array}{cccccccc}
        {1}&{}&{}&{}&{\lambda_k}&{}&{}&{} \\
        {}&{1}&{}&{}&{\lambda_{k+1}}&{}&{}&{} \\
        {}&{}&{\ddots}&{}&{\vdots}&{}&{}&{} \\
        {}&{}&{}&{1}&{\lambda_{N-1}}&{}&{}&{} \\
        {}&{}&{}&{}&{1}&{}&{} \\
        {}&{}&{}&{}&{}&{\ddots}&{} \\
        {}&{}&{}&{}&{}&{}&{1}&{} \\
        {}&{}&{}&{}&{}&{}&{}&{1} 
    \end{array}\right), \\ k=0,1,\ldots,N-1,
\end{eqnarray*}
where the empty spaces are to be understood as zero. Notice that each $A^{(k)}$ is the product of ELOs that implement Gaussian elimination on the lines of the coefficient matrix \cite{elos}.

The operation $A_1|\Phi\rangle=A_1^{(N-1)}A_1^{(N-2)}\ldots A_1^{(1)}A_1^{(0)}|\Phi\rangle$ corresponds to the left multiplication $AM_{\overline{k}}$ of each submatrix of the coefficient matrix. The operation $A_1^{(0)}|\Phi\rangle$ amounts to mapping $M_{\overline{0}}$ into
\begin{eqnarray*}
    A^{(0)}M_{\overline{0}}=\left(\begin{array}{cccccc}
         {0}&{h_1}&{h_2}&{h_3}&{\ldots}&{h_N}  \\
         {0}&{h_2}&{h_3}&{\ldots}&{h_N}&{0}  \\
         {\vdots}&{h_3}&{\vdots}&{\ \iddots}&{0}&{\vdots}  \\
         {\vdots}&{\vdots}&{h_N}&{\iddots}&{}&{}  \\
         {0}&{h_N}&{0}&{}&{}&{\vdots}  \\
         {h_N}&{0}&{\ldots}&{}&{\ldots}&{0}
    \end{array}\right)
\end{eqnarray*}
while the other submatrices are unchanged. The operation $A_1^{(1)}A_1^{(0)}|\Phi\rangle$ maps $A^{(0)}M_{\overline{0}}$ into
\begin{eqnarray*}
    A^{(1)}A^{(0)}M_{\overline{0}}=\left(\begin{array}{cccccc}
         {0}&{0}&{h_2}&{h_3}&{\ldots}&{h_N}  \\
         {0}&{0}&{h_3}&{\ldots}&{h_N}&{0}  \\
         {\vdots}&{\vdots}&{\vdots}&{\ \iddots}&{0}&{\vdots}  \\
         {\vdots}&{0}&{h_N}&{\iddots}&{}&{}  \\
         {0}&{h_N}&{0}&{}&{}&{\vdots}  \\
         {h_N}&{0}&{\ldots}&{}&{\ldots}&{0}
    \end{array}\right)
\end{eqnarray*}
while the various $M_{\overline{k}}$ with $||\vec{k}||=1$ are mapped into
\begin{eqnarray*}
    A^{(1)}M_{\overline{k}}=\left(\begin{array}{cccccc|c}
         {0}&{h_2}&{h_3}&{h_4}&{\ldots}&{h_N}&{0}  \\
         {0}&{h_3}&{h_4}&{\ldots}&{h_N}&{0}&{0}  \\
         {0}&{h_4}&{\vdots}&{\ \iddots}&{0}&{\vdots}&{\vdots}  \\
         {\vdots}&{\vdots}&{h_N}&{\iddots}&{}&{}&{\vdots}  \\
         {0}&{h_N}&{0}&{}&{}&{\vdots}&{0}  \\
         {h_N}&{0}&{\ldots}&{}&{\ldots}&{0}&{0}  \\
  \hline {0}&{0}&{\ldots}&{}&{\ldots}&{0}&{0}
    \end{array}\right)
\end{eqnarray*}
and the other submatrices are unchanged. The operation $A_1^{(2)}A_1^{(1)}A_1^{(0)}|\Phi\rangle$ maps $A^{(1)}A^{(0)}M_{\overline{0}}$ into
\begin{eqnarray*}
    A^{(2)}A^{(1)}A^{(0)}M_{\overline{0}}=\left(\begin{array}{ccccccc}
         {0}&{0}&{0}&{h_3}&{h_4}&{\ldots}&{h_N}  \\
         {0}&{0}&{0}&{h_4}&{\ldots}&{h_N}&{0}  \\
          {\vdots}&{\vdots}&{\vdots}&{\vdots}&{\iddots}&{0}&{0}\\
         {\vdots}&{\vdots}&{0}&{h_N}&{\iddots}&{0}&{\vdots}  \\
         {\vdots}&{0}&{h_N}&{0}&{\iddots}&{}&{}  \\
         {0}&{h_N}&{0}&{0}&{}&{}&{\vdots}  \\
         {h_N}&{0}&{0}&{\ldots}&{}&{\ldots}&{0}
    \end{array}\right),
\end{eqnarray*}
maps the various $A^{(1)}M_{\overline{k}}$ with $||\vec{k}||=1$ into 
\begin{eqnarray*}
    A^{(2)}A^{(1)}M_{\overline{k}}=\left(\begin{array}{cccccc|c}
         {0}&{0}&{h_3}&{h_4}&{\ldots}&{h_N}&{0}  \\
         {0}&{0}&{h_4}&{\ldots}&{h_N}&{0}&{0}  \\
         {0}&{\vdots}&{\vdots}&{\iddots}&{0}&{\vdots}&{\vdots}  \\
         {\vdots}&{0}&{h_N}&{\iddots}&{}&{}&{}  \\
         {0}&{h_N}&{0}&{}&{}&{\vdots}&{\vdots}  \\
         {h_N}&{0}&{\ldots}&{}&{\ldots}&{0}&{0}  \\
  \hline {0}&{0}&{\ldots}&{}&{\ldots}&{0}&{0}
    \end{array}\right)
\end{eqnarray*}
and the various $M_{\overline{k}}$ with $||\vec{k}||=2$ into
\begin{eqnarray*}
A^{(2)}M_{\overline{k}}=\left(\begin{array}{cccccc|cc}
         {0}&{h_3}&{h_4}&{h_5}&{\ldots}&{h_N}&{0}&{0}  \\
         {0}&{h_4}&{h_5}&{\ldots}&{h_N}&{0}&{0}&{0}  \\
         {\vdots}&{h_5}&{\vdots}&{\ \iddots}&{0}&{\vdots}&{\vdots}&{\vdots}  \\
         {\vdots}&{\vdots}&{h_N}&{\iddots}&{}&{}&{\vdots}&{\vdots}  \\
         {0}&{h_N}&{0}&{}&{}&{\vdots}&{0}&{0}  \\
         {h_N}&{0}&{\ldots}&{}&{\ldots}&{0}&{0}&{0}  \\
  \hline {0}&{0}&{\ldots}&{}&{\ldots}&{0}&{0}&{0}\\
         {0}&{0}&{\ldots}&{}&{\ldots}&{0}&{0}&{0}
    \end{array}\right)
\end{eqnarray*}
while the other submatrices are unchanged. It is easy to see then that the overall effect of $A_1|\Phi\rangle$ on the various submatrices such that $||\vec{k}||=n$ is
\begin{eqnarray*}
    M_{\overline{k}}\rightarrow\left(\begin{array}{cccccc|ccc}
         {0}&{0}&{\ldots}&{0}&{0}&{h_N}&{0}&{\ldots}&{0}  \\
         {0}&{0}&{\ldots}&{0}&{h_N}&{0}&{0}&{\ldots}&{0}  \\
         {0}&{\vdots}&{}&{\ \iddots}&{0}&{\vdots}&{\vdots}&{}&{\vdots}  \\
         {\vdots}&{0}&{\iddots}&{\iddots}&{}&{}&{}&{}&{}  \\
         {0}&{h_N}&{0}&{}&{}&{\vdots}&{\vdots}&{}&{\vdots}  \\
         {h_N}&{0}&{\ldots}&{}&{\ldots}&{0}&{0}&{\ldots}&{0}  \\
 \hline  {0}&{0}&{\ldots}&{}&{\ldots}&{0}&{0}&{\ldots}&{0}\\
         {\vdots}&{\vdots}&{}&{}&{}&{\vdots}&{\vdots}&{}&{\vdots}\\
         {0}&{0}&{\ldots}&{}&{\ldots}&{0}&{0}&{\ldots}&{0}
    \end{array}\right).
\end{eqnarray*}
Defining $S=(1/h_N)\mathcal{I}$, we see that $S_1A_1|\Phi\rangle=|\Phi_N\rangle$. $\Box$  

In Appendix B, we use Observation 3 to show that the Schmidt rank of $|\Psi\rangle$ for any bipartition of the modes is $N+1$, implying that it is a genuinely $m$-partite entangled state if $N\neq 0$. Since the Schmidt rank is invariant under SLOCC, we see that $|\Psi_n\rangle$ and $|\Psi_{n'}\rangle$ are SLOCC-inequivalent whenever $n\neq n'$. In other words, for each value of $n$ there is a SLOCC equivalence class $\mathcal{C}_n$ with representatives $|\Psi_n\rangle$ and $|\Phi_n\rangle$. The set $\mathcal{C}_n$ is convex, with mixed states comprised of convex combinations of the pure states in this SLOCC equivalence class. 

Moreover, the Schmidt rank implies the following hierarchy among the SLOCC classes \cite{sr}, as depicted in Figure 1:
\begin{eqnarray*}
    \mathcal{C}_0\subset\mathcal{C}_1\subset\mathcal{C}_2\subset\ldots\subset\mathcal{C}_{n-1}\subset\mathcal{C}_{n}\subset\mathcal{C}_{n+1}\subset\ldots
\end{eqnarray*}

Similarly to the relation between the GHZ and W classes \cite{sloccwit}, pure states in $\mathcal{C}_K$ are those SLOCC-equivalent to $|\Psi_K\rangle$. By Observation 3, all states in the form
\begin{eqnarray*}
    |\Psi\rangle=c_0|\Psi_0\rangle+c_1|\Psi_1\rangle+\ldots+c_K|\Psi_K\rangle
\end{eqnarray*}
are in $\mathcal{C}_K$ as well. Hence, we can always find a $|\Psi\rangle\in\mathcal{C}_K$ that is arbitrarily close (but never equal) to $|\Psi_k\rangle$ if $k<K$, e.g., by making $c_k\rightarrow 1$. The converse is not valid, however, since we cannot increase the Schmidt rank of any bipartition by ILOs, implying that it is not possible to find a state in $\mathcal{C}_K$ arbitrarily close to a given state in $\mathcal{C}_{K+1}$. This relation among the classes can be inferred as well from the maximum overlap between a given representative in a class and the whole set of states of some other class \cite{sloccwit2}. 

\begin{figure}[h]\centering\includegraphics[scale=0.5]{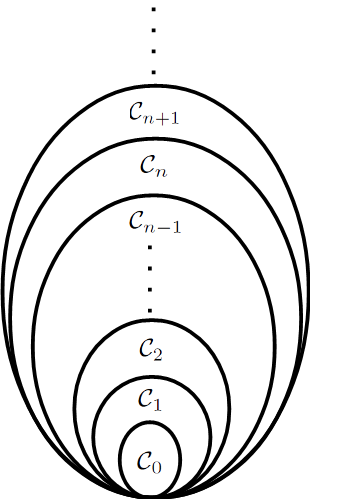}\caption{Structure of SLOCC classes for input states that are finite superpositions of number states.}
\label{fig:classesA}\end{figure}

\subsection{Superpositions of coherent states as input}

We now consider arbitrary ``cat states" \cite{cat} of the form 
\begin{eqnarray*}
    |\psi\rangle=\frac{1}{\mathcal{N}}\sum_{k=0}^{r-1}c_k|\alpha_k\rangle
\end{eqnarray*}
on the first mode, where $|\alpha_k\rangle$ are distinct coherent states and $\mathcal{N}$ is a normalization factor. The number $r$ represents the so-called nonclassicality rank \cite{nonc,nonc2} of the input state.

This finite superposition of coherent states, when going through a $m$-port beam-splitter, results in the output state
\begin{eqnarray*}
    |\Psi\rangle=U_S|\psi\rangle\otimes|0,\ldots,0\rangle=\frac{1}{\mathcal{N}}\sum_{k=0}^{r-1}c_k|\eta_{\alpha_k}\rangle
\end{eqnarray*}
where $|\eta_{\alpha_k}\rangle=|\beta_k,\ldots,\beta_k\rangle$ and $\beta_k=\alpha_k/\sqrt{m}$. 
\begin{observation}
    The output state $|\Psi\rangle$ is SLOCC-equivalent to the GHZ state
    \begin{eqnarray*}
        |GHZ_{(r)}\rangle=\frac{|0\rangle^{\otimes m}+|1\rangle^{\otimes m}+\ldots+|r-1\rangle^{\otimes m}}{\sqrt{r}}.
    \end{eqnarray*}
\end{observation}
{\it Proof:} In each mode, the states $|\beta_0\rangle,|\beta_1\rangle,\ldots,|\beta_{r-1}\rangle$ form a linearly independent set \cite{mandelwolf}. By Gram-Schmidt orthogonalization \cite{linalgebra}, there exists an invertible operator $B$ such that the states $|\chi_k\rangle=B|\beta_k\rangle$, $k=0,1,\ldots,r-1$ form an orthonormal basis. Moreover, the unitary $W=\sum_{k=0}^{r-1}|k\rangle\langle\chi_k|$ changes from the $\{|\chi_0\rangle,|\chi_1\rangle,\ldots,|\chi_{r-1}\rangle\}$ basis to the number basis $\{|0\rangle,|1\rangle,\ldots,|r-1\rangle\}$. Defining $T=\sum_{n=0}^{r-1}c_n^{-1}|n\rangle\langle n|$ and $S=(\mathcal{N}/\sqrt{r})\mathcal{I}$ we have
\begin{eqnarray*}
    S_1T_1\bigotimes_{q=1}^mW_qB_q|\Psi\rangle=\frac{1}{\sqrt{r}}\sum_{k=0}^{r-1}|k,k,\ldots,k\rangle
\end{eqnarray*}
and we have shown that $|\Psi\rangle$ is SLOCC-equivalent to the $m$-partite $GHZ$ state $|GHZ_{(r)}\rangle$. $\Box$

With a reasoning similar to the previous case, each state $|GHZ_{(r)}\rangle$ constitutes a representative of a SLOCC equivalence class $\mathcal{R}_r$, $r=1,2\ldots$. The Schmidt rank of $|GHZ_{(r)}\rangle$ for any bipartition of the state space $\mathcal{H}$ is obviously $r$. We deduce then a hierarchy based on the nonclassicality rank of the input state:
\begin{eqnarray*}
    \mathcal{R}_1\subset\mathcal{R}_2\subset\mathcal{R}_3\subset\ldots\subset\mathcal{R}_{r-1}\subset\mathcal{R}_r\subset\mathcal{R}_{r+1}\subset\ldots
\end{eqnarray*}

\begin{figure}[h]\centering\includegraphics[scale=0.5]{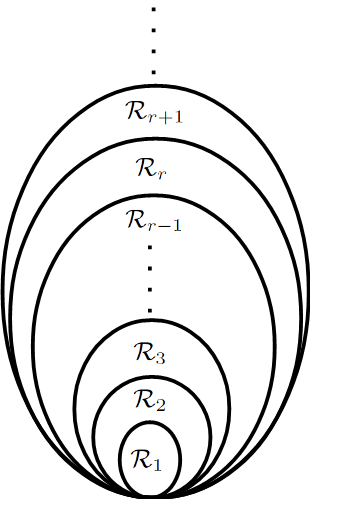}\caption{Structure of SLOCC classes for input states that are finite superpositions of coherent states.}
\label{fig:classesB}\end{figure}

Let $C=\sum_{n=0}^{r-1}d_n|n\rangle\langle n|$. Then we have  
\begin{eqnarray*}
   C_1|GHZ_{(r)}\rangle=\frac{1}{\sqrt{r}}\sum_{k=0}^{r-1}d_k|k,k,\ldots,k\rangle
\end{eqnarray*}
It is clear then that $|GHZ_{(r)}\rangle$ can be made arbitrarily close to $|GHZ_{(r')}\rangle$ for $r'<r$, but the converse is not true for $r'>r$. 

\subsection{Hybrid superpositions}

A more exotic example of input state is a finite superposition of both number and coherent states,
\begin{eqnarray*}
    |\psi\rangle=\frac{1}{\mathcal{N}}\left(\sum_{n=0}^Nc_n|n\rangle+\sum_{k=0}^{r-1}d_k|\alpha_k\rangle\right),
\end{eqnarray*}
where $\mathcal{N}$ stands for a normalization factor. The output state produced by a MBS when $|\psi\rangle$ is the input is, according to the previous discussion:
\begin{eqnarray*}
    |\Psi\rangle=\frac{1}{\mathcal{N}}\left(\sum_{n=0}^Nc_n|\Psi_n\rangle+\sum_{k=0}^{r-1}d_k|\eta_{\alpha_k}\rangle\right),
\end{eqnarray*}
where $|\eta_{\alpha_k}\rangle=|\beta_k,\ldots,\beta_k\rangle$ and $\beta_k=\alpha_k/\sqrt{m}$. 
\begin{observation}
    The output state $|\Psi\rangle$ is SLOCC-equivalent to the state
\begin{eqnarray*}
    |\Phi_N\rangle+\sum_{n=N+1}^{N+r}|n\rangle^{\otimes m}.
\end{eqnarray*}
\end{observation}
{\it Proof:} The set of states $\{|0\rangle,|1\rangle,\ldots,|N\rangle\}\cup\{|\beta_0\rangle,|\beta_1\rangle,\ldots,|\beta_{r-1}\rangle\}$ is linearly independent in each mode. By Gram-Schmidt orthogonalization \cite{linalgebra}, there exists an invertible operator $F$ such that $F|n\rangle=|n\rangle$, $n=0,1,\ldots,N$ and $|\mu_k\rangle=F|\beta_k\rangle$, $k=0,1,\ldots,r-1$, and such that $\{|0\rangle,|1\rangle,\ldots,|N\rangle\}\cup\{|\mu_0\rangle,|\mu_1\rangle,\ldots,|\mu_{r-1}\rangle\}$ form an orthonormal basis. The unitary $V=\sum_{n=0}^N|n\rangle\langle n|+\sum_{k=N+1}^{N+r}|k\rangle\langle\mu_{k-(N+1)}|$ changes from the $\{|0\rangle,|1\rangle,\ldots,|N\rangle\}\cup\{|\mu_0\rangle,|\mu_1\rangle,\ldots,|\mu_{r-1}\rangle\}$ basis to the number basis $\{|0\rangle,|1\rangle,\ldots,|N\rangle,|N+1\rangle,|N+2\rangle,\ldots,|N+r\rangle\}$. Defining $Q=\sum_{n=0}^N|n\rangle\langle n|+\sum_{n=N+1}^{N+r}d_n^{-1}|n\rangle\langle n|$ and $P=(\mathcal{N})\mathcal{I}$ we have
\begin{eqnarray*}
    Q_1P_1\bigotimes_{q=1}^mV_qF_q|\Psi\rangle=\sum_{n=0}^Nc_n|\Psi_n\rangle+\sum_{n=N+1}^{N+r}|n\rangle^{\otimes m}.
\end{eqnarray*}
Let $G=\sum_{n=0}^N\sqrt{n!}|n\rangle\langle n|+\sum_{n=0}^N|n\rangle\langle n|$; we have, according to the proof of Observation 2:
\begin{eqnarray*}
   \bigotimes_{q=1}^mG_q\left(\sum_{n=0}^Nc_n|\Psi_n\rangle+\sum_{n=N+1}^{N+r}|n\rangle^{\otimes m}\right)\\ =\sum_{n=0}^Nh_n|\Phi_n\rangle+\sum_{n=N+1}^{N+r}|n\rangle^{\otimes m}.
\end{eqnarray*}
Let $A$ and $S$ be the operators used in the proof of Observation 3. Then we have
\begin{eqnarray*}
    S_1A_1\left(\sum_{n=0}^Nh_n|\Phi_n\rangle+\sum_{n=N+1}^{N+r}|n\rangle^{\otimes m}\right)=|\Phi_N\rangle+\sum_{n=N+1}^{N+r}|n\rangle^{\otimes m}
\end{eqnarray*}
and the output state $|\Psi\rangle$ is thus SLOCC-equivalent to $|\Phi_N\rangle+\sum_{n=N+1}^{N+r}|n\rangle^{\otimes m}$. $\Box$

By previous results, the Schmidt rank of $|\Phi_N\rangle+\sum_{n=N+1}^{N+r}|n\rangle^{\otimes m}$ is $N+r+1$. Hence, there are different values of $N$ and $r$ where the respective states states have equal Schmidt ranks. For example, the states $|\Phi_3\rangle+\sum_{n=4}^{5}|n\rangle^{\otimes m}$ and $|\Phi_2\rangle+\sum_{n=3}^{5}|n\rangle^{\otimes m}$ both have Schmidt rank $6$. As discussed in the next session, these states are SLOCC-inequivalent, despite having equal Schmidt ranks. 

\section{Comparison between the three scenarios}

We considered the three types of superpositions of the previous section separately and, in order to determine whether each type of representative is equivalent or inequivalent under SLOCC, we invoke the following result from \cite{chenchen,chenchenmei}:
\begin{theorem}
Two pure states of a multipartite system are equivalent under SLOCC if and only if (i) they have the same local rank of each party, and (ii) the
ranges of the adjoint reduced density matrices of each party of them are related by certain ILO's.
\end{theorem}
By adjoint reduced density matrices it is understood the density matrices obtained through discarding (partial trace) of one of the local subsystems $\mathcal{H}_k$ of the global system $\mathcal{H}=\mathcal{H}_1\otimes\mathcal{H}_2\otimes\ldots\otimes\mathcal{H}_m$. Denoting by $a_k$ the number of product states in the range of the adjoint reduced density matrix related to $\mathcal{H}_k$, it is a consequence of Theorem 1 above that if two states are SLOCC-equivalent, then the array of values $[a_1,a_2,\ldots,a_m]$ of these states must be equal. Or, equivalently, if any value $a_k$ differs for these states, then they are SLOCC-inequivalent. 

Due to the full permutational symmetry of the states considered and the simple form of the representatives in each scenario, the values $a_k$ of a given state are all equal and are easily obtained. Following the notation in \cite{chenchen,chenchenmei}, we have the following classification:
\begin{itemize}
    \item $|\Psi_N\rangle\sim|\Phi_N\rangle\in [1,1,\ldots,1]$.
    \item $|GHZ_{(r)}\rangle\in[r,r,\ldots,r]$.
    \item $|\Phi_N\rangle+\sum_{n=N+1}^{N+r}|n\rangle^{\otimes m}\in[r+1,r+1,\ldots,r+1]$.
\end{itemize}
It is straightforward then that each representative is SLOCC-inequivalent to the other, even if they have the same Schmidt rank. For example, the Schmidt rank of $|GHZ_{(M)}\rangle$, $|\Psi_{M-1}\rangle$ and $|\Phi_{M-2}\rangle+|M-1\rangle^{\otimes m}$ is $M$, but each is respectively in $[M,M,\ldots,M]$, $[1,1,\ldots,1]$ and $[2,2,\ldots,2]$, being thus inequivalent by SLOCC.    

The inequivalence between the three scenarios does not imply a hierarchy among the different SLOCC classes. Only in the multiqubit situation there are analytical results concerning the relationship between the multiqubit GHZ and W classes \cite{sloccwit,sloccwit2}, coming from parametrizations of the symmetric subspaces of multiqubits. The hierarchies between the SLOCC classes described here will thus remain an open question.  

\section{Conclusions and perspectives}

In this manuscript, we investigated the different multipartite entanglement classes that arise in a multiport beam-splitter in three distinct situations, depending on the single-mode states that enter the device. In the first scenario, we considered input states that are finite superpositions of number states and concluded that the different SLOCC equivalence classes were related to the highest total number of the output state. Physically, this is the highest energy attained by the state and thus more energy results in more ``powerful" multipartite entanglement. In the second scenario, we analysed input states that are finite superpositions of coherent states and identified a hierarchy among the SLOCC classes, but this time in terms of the nonclassicality rank of the input state. In the third scenario, we considered a hybrid situation where the input states are a superposition of both number and coherent states, obtaining a SLOCC classification that is a combination of the other two scenarios. The multipartite entanglement classes corresponding to each scenario were shown to be all inequivalent to each other. 

The results obtained here give possible alternatives for the generation of target multipartite entangled states with desirable properties. In regards to entanglement properties related to the SLOCC equivalence class of a state, our work shows that one could employ as an input a finite superposition of number states - which can be obtained by truncation of a suitable continuous-variable state \cite{truncation} - instead of using a number state itself, whose implementation can be challenging. Likewise, instead of generating a GHZ-like state, whose implementation can be even more challenging, by using cat states - which can be generated in various experimental proposals in the literature - as inputs on a MBS, one obtains a multipartite entangled state with the same SLOCC properties. Moreover, there are limitations to the creation of certain types of states in linear quantum optics \cite{nogo} and thus the possibility of finding alternatives should be pursued.       

For the classification of mixed states, one could employ the techniques in \cite{sloccwit,sloccwit2} in order to detect to which multipartite entanglement classes a given mixed output state belongs, besides the use of the Schmidt number \cite{sr}, which is the mixed state analogue of the Schmidt rank. A related problem is the physical implementation of so-called SLOCC witnessses in the quantum optical domain. These interesting and complex open questions go beyond the scope of the present work and will be investigated elsewhere.

\section{Acknowledgments}

This work is part of the institutional research projects $376/2020$ ``Characterization of quantum correlations" and $475/2023$ ``Mathematical aspects of quantum entanglement" from Universidade Federal de Mato Grosso. The author acknowledges the technical support of the Wolfram Research team and is thankful to Jo\~ao Bosco de Siqueira for ideas and discussions. 

\begin{widetext}

\section{Appendix} 

\subsection{Generality of a balanced multiport beam-splitter}

An arbitrary state on the first mode is given by 
\begin{eqnarray*}
    |\psi\rangle=\sum_{n=0}^{\infty}c_n|n\rangle=\sum_{n=0}^{\infty}c_n\frac{(a_1^{\dagger})^n}{\sqrt{n!}}|0\rangle
\end{eqnarray*}
If this state goes through an arbitrary $m$-port beam-splitter, the resulting output state is given by 
\begin{eqnarray*}
    |\tilde{\Psi}\rangle &=& \sum_{n=0}^{\infty}\frac{c_n}{\sqrt{n!}}(\gamma_1a_1^{\dagger}+\gamma_2a_2^{\dagger}+\ldots+\gamma_ma_m^{\dagger})^n|0,0,\ldots,0\rangle \\
    &=&\sum_{n=0}^{\infty}\frac{c_n}{\sqrt{n!}}\sum_{n_1+\ldots+n_m=n}{n\choose n_1,n_2,\ldots,n_m}(\gamma_1a_1^{\dagger})^{n_1}(\gamma_2a_2^{\dagger})^{n_2}\ldots(\gamma_ma_m^{\dagger})^{n_m}|0,0,\ldots,0\rangle\\
     &=&\sum_{n=0}^{\infty}c_n\sum_{n_1+\ldots+n_m=n}\sqrt{n\choose n_1,n_2,\ldots,n_m}\gamma_1^{n_1}\gamma_2^{n_2}\ldots\gamma_m^{n_m}|n_1,n_2,\ldots,n_m\rangle
\end{eqnarray*}
where $\gamma_k\in\mathbb{C}$ and $|\gamma_1|^2+|\gamma_2|^2+\ldots+|\gamma_m|^2=1$. Defining the following (bounded) ILO on mode $q$,
\begin{eqnarray*}
    D_q=\frac{1}{\sqrt{m}}\sum_{q=0}^{\infty}(\gamma_q^{-1})^{n_q}|n_q\rangle\langle n_q|
\end{eqnarray*}
we see that its effect on the output state is
\begin{eqnarray*}
    \bigotimes_{q=1}^{m}D_q|\tilde{\Psi}\rangle &=& \sum_{n=0}^{\infty}c_n\sum_{n_1+\ldots+n_m=n}\sqrt{n\choose n_1,n_2,\ldots,n_m}\gamma_1^{n_1}\gamma_2^{n_2}\ldots\gamma_m^{n_m}\bigotimes_{q=1}^{m}D_q|n_1,n_2,\ldots,n_m\rangle \\
    &=&\sum_{n=0}^{\infty}\frac{c_n}{m^{n/2}}\sum_{n_1+\ldots+n_m=n}\sqrt{n\choose n_1,n_2,\ldots,n_m}|n_1,n_2,\ldots,n_m\rangle,
\end{eqnarray*}
which is the expression for the output state of a balanced MBS.

\subsection{Schmidt rank of $|\Phi_N\rangle$}

Without loss of generality, let us consider a bipartition $(\mathcal{H}_1\otimes\mathcal{H}_2\otimes\ldots\otimes\mathcal{H}_{k-1})\otimes(\mathcal{H}_k\otimes\mathcal{H}_{k+1}\otimes\ldots\otimes\mathcal{H}_m)$ of the global state space $\mathcal{H}$. The Schmidt decomposition of the state is given by
\begin{eqnarray*}
    |\Phi_N\rangle&=& |00\ldots 00\rangle\otimes\left(\sum_{n_k+n_{k+1}+\ldots+n_m=N}|n_k,n_{k+1},\ldots,n_m\rangle\right)\\
    &+&\left(\sum_{n_1+n_2+\ldots+n_{k-1}=1}|n_1,n_2,\ldots,n_{k-1}\rangle\right)\otimes\left(\sum_{n_k+n_{k+1}+\ldots+n_m=N-1}|n_k,n_{k+1},\ldots,n_m\rangle\right)\\
    &+&\left(\sum_{n_1+n_2+\ldots+n_{k-1}=2}|n_1,n_2,\ldots,n_{k-1}\rangle\right)\otimes\left(\sum_{n_k+n_{k+1}+\ldots+n_m=N-2}|n_k,n_{k+1},\ldots,n_m\rangle\right)\\
    &+&\ldots +\left(\sum_{n_1+n_2+\ldots+n_{k-1}=N}|n_1,n_2,\ldots,n_{k-1}\rangle\right)\otimes|00\ldots 00\rangle
\end{eqnarray*}
We conclude that the Schmidt rank of $|\Phi_N\rangle$ is $N+1$.

\end{widetext}


\begin{thebibliography}{99}

\bibitem{nielsen} M. A. Nielsen and I. L. Chuang (2000), {\it Quantum computation and quantum information},
Cambridge University Press (Cambridge).

\bibitem{multientang} R. Horodecki, P. Horodecki, M. Horodecki and K. Horodecki, Rev. Mod. Phys. {\bf 81},  865 (2009).

\bibitem{cris} C. Ritz, {\it  Characterizing the structure of multiparticle entanglement in high-dimensional systems }, PhD Thesis, University of Siegen (2019). 

\bibitem{gtreview} O. G\"uhne and G. T\'oth, Phys. Rep. \textbf{474}, 1 (2009).

\bibitem{manybody} L. Amico, R. Fazio, A. Osterloh, V. Vedral, Rev. Mod. Phys. {\bf 80}, 517.

\bibitem{mps} M. Sanz, I. L. Egusquiza, R. Di Candia, H. Saberi, L. Lamata, E. Solano, Sci. Rep. {\bf 6}.

\bibitem{nonlocality}  	A. Cabello, Phys. Rev. A {\bf 65} 032108 (2002).

\bibitem{wghz} W. D\"ur, G. Vidal, J. I. Cirac, Phys. Rev. A {\bf 62}, 062314 (2000).

\bibitem{miyake} A. Miyake, Phys. Rev. A 67, 012108 (2003).

\bibitem{verstraete} F. Verstraete, J. Dehaene, B. De Moor, Phys. Rev. A 68, 012103 (2003). 

\bibitem{lamata} L. Lamata, J. Leon, D. Salgado, and E. Solano, Phys. Rev. A {\bf 74}, 052336 (2006).

\bibitem{dafa} X. Li, D. Li, Phys. Rev. Lett. {\bf 108}, 180502 (2012).

\bibitem{gour} G. Gour and N. R. Wallach, Phys. Rev. Lett. 111, 060502 (2013).

\bibitem{elos} F. E. S. Steinhoff, Phys. Rev. A 100, 022317 (2019).

\bibitem{mbs} S. Kumar, D. Bhatti, A. E. Jones, S. Barz, New J. of Phys. 25, 063027 (2023).

\bibitem{mbs2} D. Bhatti, S. Barz, Phys. Rev. A 107, 033714 (2023).

\bibitem{nonc} N. Killoran, F.E.S. Steinhoff, M.B. Plenio Phys. Rev. Lett. {\bf 116}, 080402 (2016).

\bibitem{nonc2} B. Regula, M. Piani, M. Cianciaruso, T. R. Bromley, A. Streltsov, G. Adesso	New J. Phys. {\bf 20}, 033012 (2018). 

\bibitem{nonc3} J. K. Asb\'oth, J. Calsamiglia, H. Ritsch, Phys. Rev. Lett. 94, 173602 (2005).

\bibitem{nonc4} J. Solomon Ivan, S. Chaturvedi, E. Ercolessi, G. Marmo, G. Morandi, N. Mukunda, R. Simon, Phys. Rev. A 83, 032118 (2011).


\bibitem{infinite} M. Owari, K. Matsumoto, M. Murao, Phys. Rev. A 70, 050301(R) (2004).

\bibitem{ds1} Z. Li, Y.-G. Han, H.-F. Sun, J. Shang, H. Zhu, Phys. Rev. A 103, 022601 (2021).

\bibitem{ds2} R.I. Nepomechie, D. Raveh, arXiv:2301.04989 

\bibitem{ds3} J. Romero-Pallej\`a, J. Ahiable, A. Romancino, C. Marconi, A. Sanpera, arXiv:2403.05244 

\bibitem{ds4} R.I. Nepomechie, F. Ravanini, D. Raveh,  	arXiv:2402.03233
 




\bibitem{lf} N. Gisin, Phys. Lett. A {\bf 210}, 151 (1996).

\bibitem{lf2} J. Bowles, J. Francfort, M. Fillettaz, F. Hirsch, N. Brunner, Phys. Rev. Lett. {\bf 116}, 130401 (2016).

\bibitem{vidal} G. Vidal, Phys. Rev. Lett. 83, 1046 (1999). 


















\bibitem{puri} R.R. Puri {\it Mathematical Methods of Quantum Optics}, Springer Science and Business Media (2001).

\bibitem{futurework} F. E. S. Steinhoff {\it et al} (work in preparation).

\bibitem{goldbergjames} A.Z. Goldberg, D.F.V. James, J. Phys. A: Math. Theor. 51, 385303 (2018).

\bibitem{multentanggen} M.C. Tichy, F. Mintert, A. Buchleitner, Phys. Rev. A 87, 022319 (2013).

\bibitem{cm} Y. Huang, H. Yu, F. Miao, T. Han, X. Zhang, Int. J. Quant. Inf. 20, 2150035 (2022).

\bibitem{cm2} J. Chang., N. Jing, T. Zhang, Int J Theor Phys 62, 6 (2023). 

\bibitem{cm3} S. Wang, Y. Lu, M. Gao, J. Cui, J. Li, J. Phys. A: Math. Theor. 46 105303 (2013). 

 

\bibitem{truncation} M. Koniorczyk, Z. Kurucz, A. G\'abris, and J. Janszky, Phys. Rev. A 62, 013802 (2000).

\bibitem{truncation2} W. Leo\'nski, A. Kowalewska-Kudlaszyk, Progress in Optics 56, 131 (2011). 

\bibitem{cutoff} J. J. Guanzon, M. S. Winnel, A. P. Lund, T. C. Ralph, Phys. Rev. Lett. 128, 160501 (2022). 


\bibitem{becv} F. E. S. Steinhoff, M. C. de Oliveira, J. Sperling, W. Vogel, Phys. Rev. A {\bf 89}, 032313 (2014).

\bibitem{horn} R. A. Horn and C. R. Johnson, Matrix Analysis (Cambridge University Press, Cambridge, 1985).

















\bibitem{sr} C. Zhang, S. Denker, A. Asadian, O. G\"uhne, arXiv:2304.02447.

\bibitem{sloccwit} A. A\'cin, D. Bruss, M. Lewenstein, A. Sanpera, Phys. Rev. Lett. 87, 040401 (2001).

\bibitem{sloccwit2} C. Ritz, C. Spee, O. G\"uhne, J. Phys. A: Math. Theor. 52, 335302 (2019).

\bibitem{mandelwolf} L. Mandel, E. Wolf, {\it Optical coherence and quantum optics}, Cambridge University Press (1995).


\bibitem{linalgebra} K. Hoffman and R. Kunze, Linear Algebra (Prentice-Hall Mathematics Series, Upper Saddle River, 1962).

\bibitem{cat} B. Vlastakis {\it et al}, 
Science 342, 607  (2013).

\bibitem{chenchen} Lin Chen, Yi-Xin Chen, Phys. Rev. A 73, 052310 (2006).

\bibitem{chenchenmei} Lin Chen, Yi-Xin Chen, Yu-Xue Mei, Phys. Rev. A 74, 052331 (2006).

\bibitem{nogo} P.V. Parellada, V.G. Garcia, J.-J. Moyano-Fern\'andez, J.C. Garcia-Escartin, arXiv:2307.11478.

\end{thebibliography}
\end{document}